\documentstyle[11pt,newpasp,twoside,epsf]{article}
\begin{document}
\title{Galactic Structure and Evolution: a decade of surveys
}
 \author{Gerry Gilmore}
\affil{Institute of Astronomy, Madingley Rd., Cambridge CB3 0HA, UK}

\begin{abstract}

Surveys of the local and distant Universe are the means to test and
improve our models of galaxy formation. Substantial successes in the
models are evident, while there is also considerable recent progress
in identifying what remains to be learned. The key weaknesses of
present models are related to merging histories, and small-scale
structures, which are both significantly at variance with
observations.  Observers are polite, and often emphasise their
agreements with models.  data is objective, and shows us the way to
focus future surveys, to allow improved understanding and knowledge.

\end{abstract}

\section{Introduction}

The mass assembly history of galaxies remains one of the critical
issues in observational cosmology: did galaxies reach their present
stellar mass only recently (say, at $z\sim 1$)? Or were most
(massive) galaxies already in place by $z\sim 1$? Is late/current
accretion significant or merely a perturbation? 
These questions may best be addressed by a combination of detailed
surveys of the Local Group, where quantitative resolved information
may be obtained, together with spectroscopic surveys of galaxies as a
function of redshift. Substantial progress on both fronts is being
made through the wealth of surveys underway and recently reported.

Surveys are indeed the currently favoured means to progress in
astrophysics: a search through the journals or astro-ph for `survey'
immediately overloads any reading list. Equally noticeable is the
current fashion to add a speculation about current mergers to any
paper studying any object in the Local Group, irrespective of its
direct relevance to the specific data at hand. More fundamental is to
consider if the rate and amplitude of the (clearly still happening)
current accretion into the Milky Way is signal or noise, and is or is
not consistent with model expectations.  

That is, we should not aspire to show that current formation models
correctly predict the Local Group, and its current merger activity: we
know, from direct observation, that current models require many better
constraints to allow their improvement, to correspond to real
galaxies. Our goal is to consider the observations objectively, to
quantify those constraints.

Spectroscopic surveys of faint galaxies, especially those selected in
the $K$-band, currently offer an excellent opportunity to address these
questions for the distant universe (Broadhurst et al. 1992; Cimatti
etal 2003: see especially {\tt
http://www.arcetri.astro.it/$\sim$k20/}).  The results of these
surveys, together with current simulations of galaxy formation,
provide the context in which Local Group studies can be interpreted,
by highlighting those aspects of current models where most progress is
required. 

One specific result of current deep surveys, which has been
long established in the Galaxy, is the apparently long time since the
galaxies were assembled. This result, long known locally, and the
subject of many continuing local investigations (eg Sandage, Lubin and
VandenBerg 2003; see also N\"ordstrom etal 2004) identifies one of the
most promising ways in which our understanding of galaxy formation can
be improved and extended. In essence, galaxies seem to have formed
earlier than current models expect, and have remained (relatively)
undisturbed for a longer time than predicted. Extending and
explaining this contradiction promises to improve significantly our
understanding not only of galaxy evolution, but also of the coupled
question, the distribution of Dark Matter on small scales, and its
corresponding temperature.

In one of the most impressive of recent studies, Cimatti etal (2002)
present the redshift distribution of a complete sample of 480 galaxies
with $K_s<20$ distributed over two independent fields covering a total
area of 52 arcmin$^2$.  A ``blind'' comparison was made with the
predictions of a set of the most recent $\Lambda$CDM hierarchical
merging (HMM) and a set of pure luminosity evolution (PLE) models. The
hierarchical merging models overpredict and underpredict the number of
galaxies at low-$z$ and high-$z$ respectively, whereas the PLE models
match the median redshift and the low-$z$ distribution, while still
being able to follow the high-$z$ tail of $N(z)$. These results are
summarised in Figure 1.

\begin{figure}
\plottwo{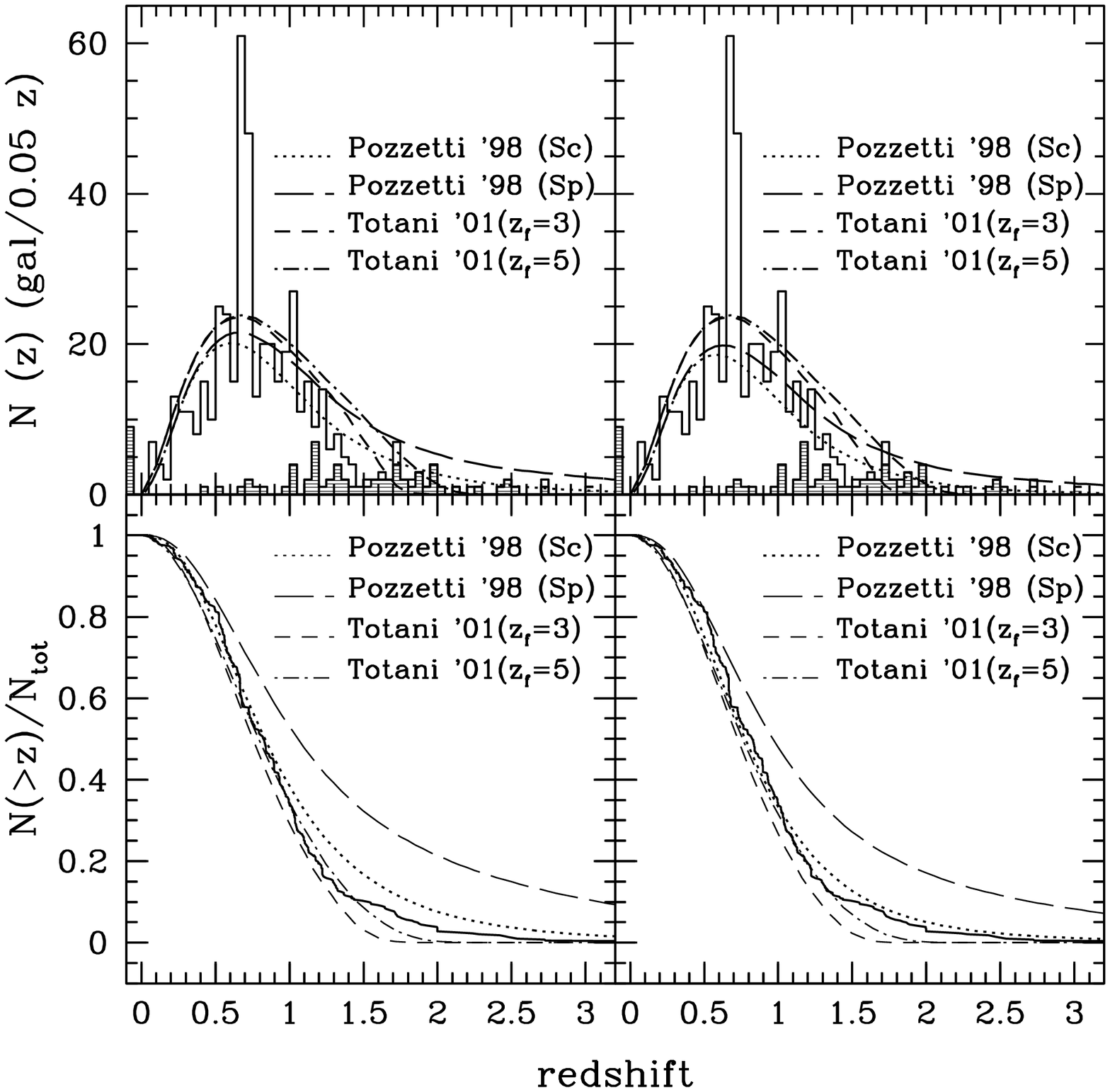}{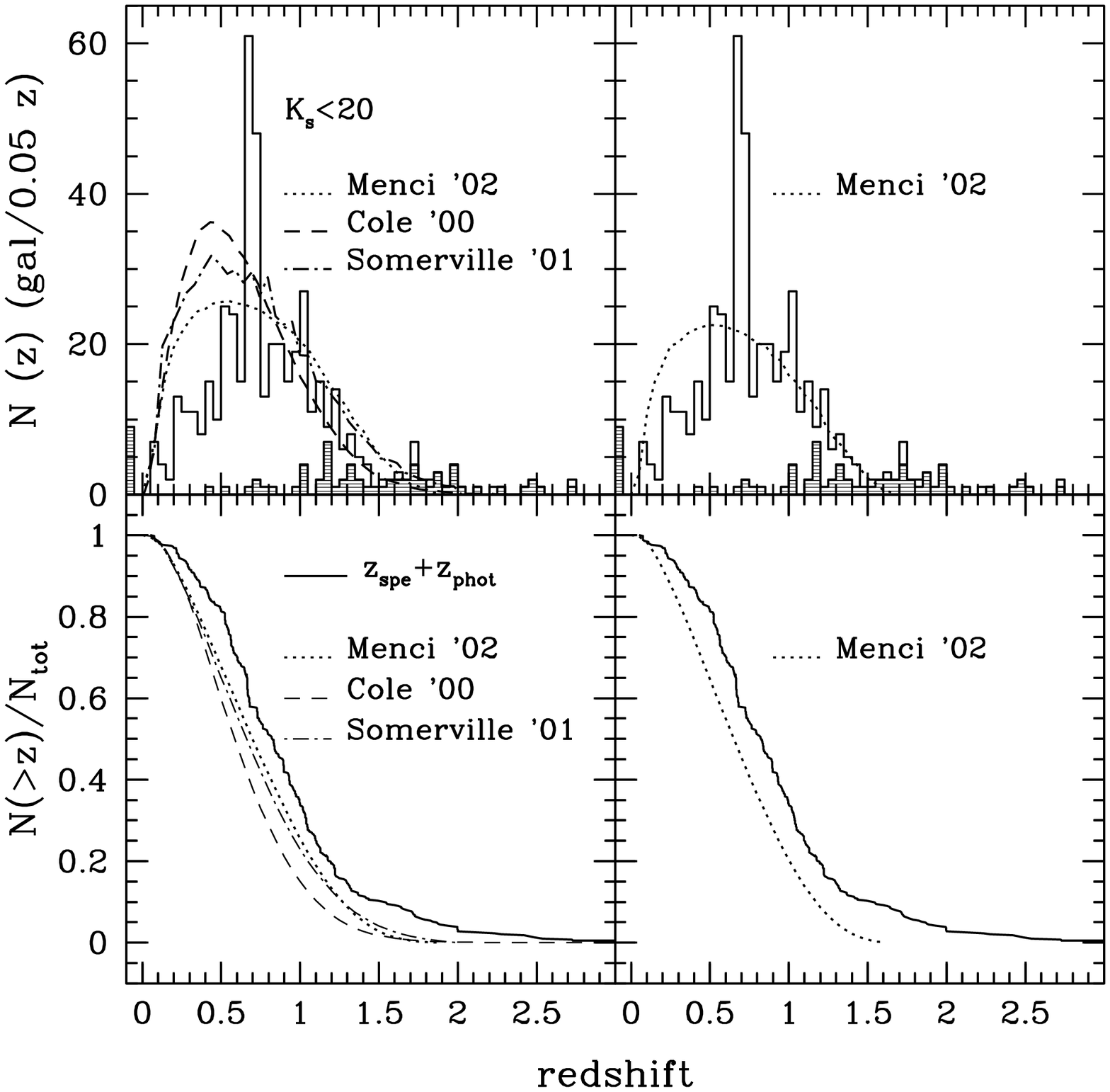}
\caption{
{\it LHS: Top panels:} the observed differential $N(z)$ for 
$K_s<20$ (histogram) compared with the PLE model predictions. 
{\it LHS Bottom panels:} the observed fractional cumulative redshift 
distribution (continuous line) compared with the same models.
The {\it left} and {\it right} panels show the models without and with the 
inclusion of the photometric selection effects respectively. Sc and Sp
indicate Scalo and Salpeter IMFs respectively.
{\it RHS: Top panels:} the observed differential redshift distribution for 
$K_s<20$ (histogram) compared with the HMM predictions. {\it RHS: Bottom
panels:} the observed fractional cumulative redshift distribution 
(continuous line) compared with the same models of top panels.
The {\it right} panels show the  model with the inclusion of the 
photometric selection effects.
These figures and captions are from Cimatti etal 2002. 
}
\end{figure}

These observational surveys illustrate the areas where current
paradigmatic models can be most improved. In particular, $\Lambda$CDM
hierarchical merging models as they are currently implemented
overpredict the total number of galaxies with $K_s<20$ in the K20
survey area by factors of 30-45\%, and are inconsistent with the
observed galaxy redshift distribution by substantial factors,
especially at high redshifts. A Kolmogorov-Smirnov test showed that
all the hierarchical merging models disagree with the observations at
$>99$\% level.

Nonetheless, the hierarchical merging $\Lambda$CDM scenario has
spectacular success in reproducing large scale structure and CMB
results, and is certainly the paradigm of choice in which we should
analyse survey and galactic structure data today. Our ambition is to
improve the ability of this framework to become consistent with the
real physics of small-scale structure and galaxy formation, by
identifying areas in which the input physics can be improved. The
observed discrepancies derived from the high-redshift studies are
related to what is essentially a purely heuristic set of algorithms
adopted to describe the star formation processes and their feedback,
both within individual galaxies and in their environment. The
high-redshift results suggest that the galaxy formation recipes should
allow galaxy formation in a CDM dominated universe to occur in a
manner described by the old-fashioned pre-CDM {\it monolithic
collapse} scenario.  This requires the recipes to allow enhanced
merging and star formation in massive haloes at high redshift (say,
$z\sim 3$), while at the same time suppressing star formation in
low-mass haloes.

We now ask if this improved recipe would be consistent with the Milky
Way, an L$^*$ galaxy whose properties should not be too rare in any
plausible formation scenario.

\section{ Mergers, phase-space structures, fashion
accessories}

Search for and interpretation of phase-space structure in the Galaxy
has always been important, but has recently become a major
industry. Significant structure is unexpected in a context where
galaxy photometrists provid convincing evidence for symmetric and
smooth surface brightness profiles (eg Kormendy \& Djorgovski 1989),
where density profiles for special tracers (RR Lyraes being popular)
show smooth profiles over very many scale lengths, and where local
stellar kinematics looks closely Gaussian (cf Gilmore, Wyse \& Kuijken
1989): clearly whatever process makes galaxies makes smooth-looking
relaxed-looking galaxies. Nonetheless, testing the paradigm is
important. The current fashion is perhaps to start from the other
extreme (cf Freeman \& Bland-hawthorn 2002) where some aspire to
``associate components of the Galaxy to elements of the protocloud'',
an interesting alternative to the statistical distribution function
analyses on which the science of stellar dynamics has been built.
 
Earlier searches for structure in the stellar distribution
used techniques such as correlation functions (eg Gilmore, G.; Reid, N.;
Hewett, P. 1985) or presumed local evolutionary processes were the
default explanation of identified field halo clustering (eg Arnold \&
Gilmore 1992). The many studies of Eggen on moving groups were perhaps
not given due recognition, since most were of thin disk stars,
disk structure is inevitably lumpy in phase space, and tells little of
global or long-term galaxy evolution. His study of the Arcturus
Moving Group was however long understood as significant for thick-disk
studies (cf Fuhrmann 2004, sect 2.4) and appears as such in
encyclopeadias, though is still being rediscovered. [Eggen however (private
communication) was deeply sceptical of the distinction between thin and
thick disk moving groups.] gould (2003) has recently reminded us that
an apparently smooth distribution is either really smooth - in which
case its history is lost into statistical distribution functions -- or
the central limit theorem applies -- in which
case its history is lost into statistical distribution functions.

The field changed rapidly after discovery of the Sgr
dwarf galaxy (Ibata, Gilmore \& Irwin 1994, 1995), an object genuinely
discovered in phase-space in real-time at the AAT where Ibata and
Gilmore were plotting colour-magnitude-velocity diagrams during their
study of the Galactic Bulge (Ibata \& Gilmore 1995a, 1995b). 
While this discovery certainly activated the field (though the
forthcoming 2MASS and SDSS surveys would have made the same discovery
through traditional photometric, rather than phase-space, techniques)
the reson it was possible seems not to have been fully appreciated by
those attempting to emulate the discovery: Sgr was found because its
stellar population is very different from that of the Galactic bulge
or the Galactic halo. It is an intermediate age and intermediate
metallicity population, and is unique in that respect.

One (should have) immediately deduced that Sgr was a rare event, not a
paradigm for the average (see Unavane, Wyse \& Gilmore 1996). The
outcome of a truly impressive and truly vast amount of work since its
discovery mapping the distribution of Sgr-related debris trails around
the Galaxy has reinforced this early conclusion: while much of Sgr is
strewn around the Galaxy,  nothing else like it has left any trace of
its demise in the last Hubble time. Except, of course, the Thick Disk
(Gilmore \& Reid 1983), whenever that got into place.

\begin{figure}[ht]
\plottwo{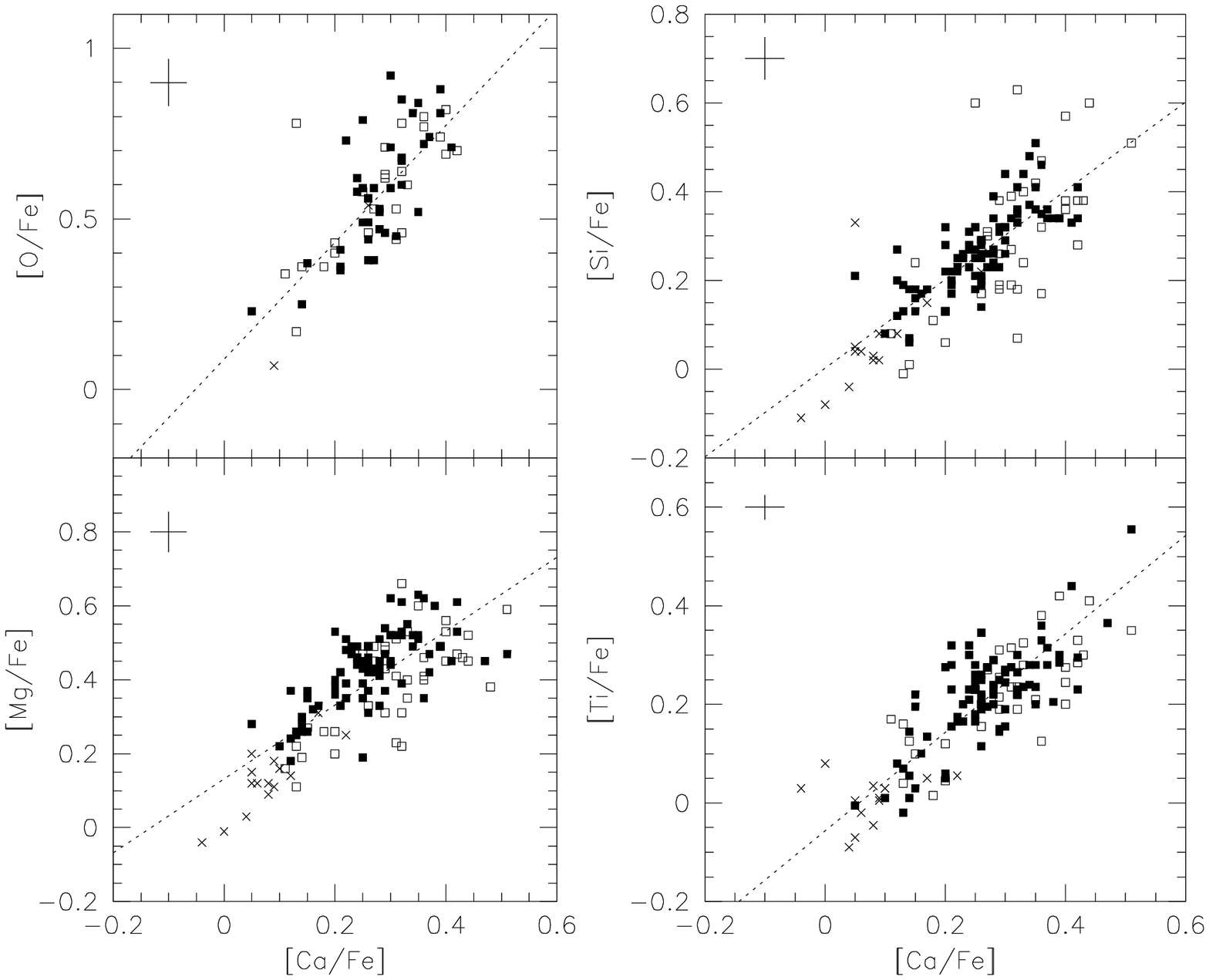}{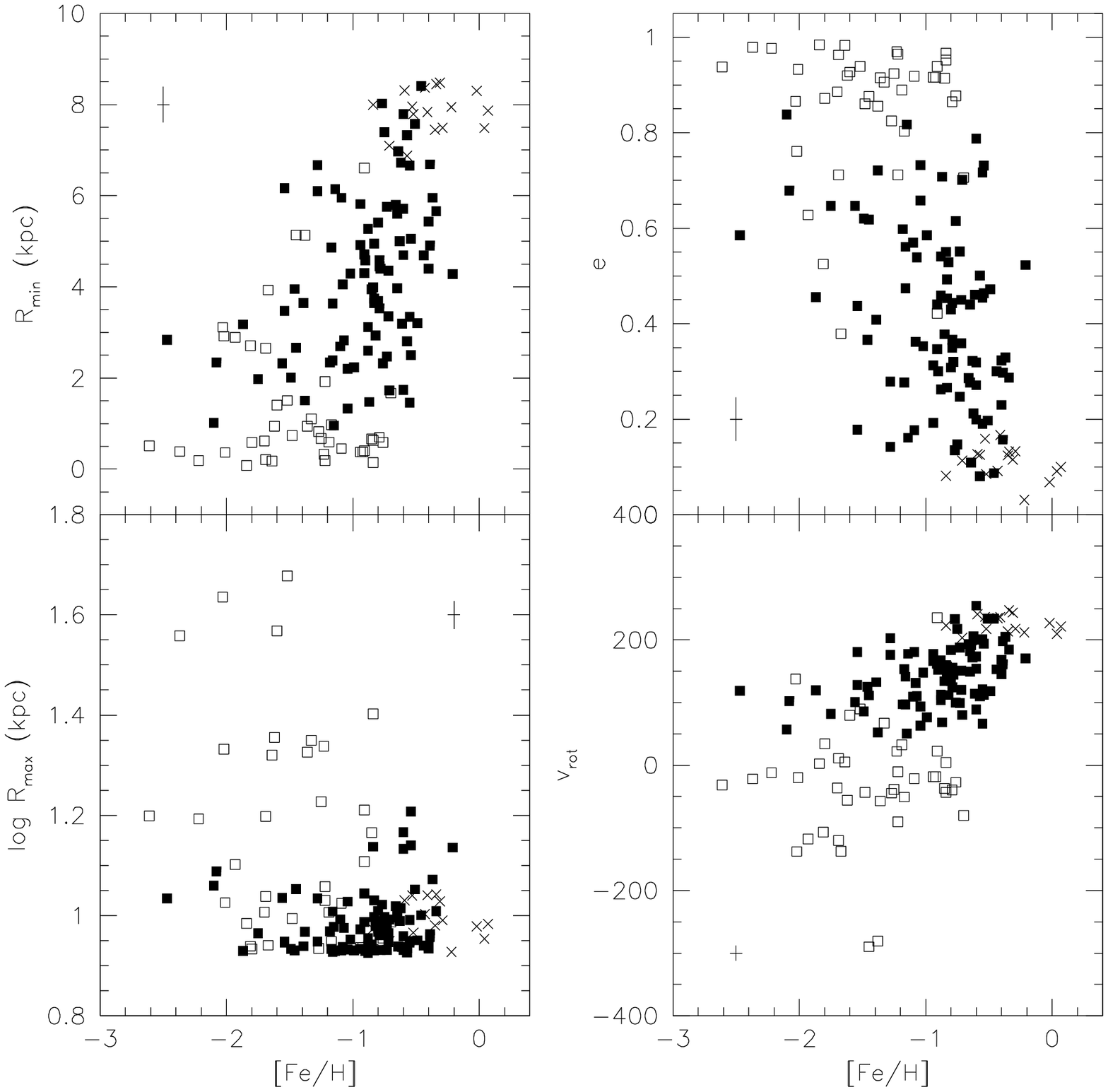}
\caption{Element ratio data (LHS) and kinematic correlations (RHS) in
metal poor stars in the Galactic halo. These figures are from Gratton
etal 2003.}
\end{figure}

\section{The chemical elements as archaeological tracers}

The recent surveys which have delivered the most quantitative
information on Galactic history are those which provide high-precision
elemental abundances for many stars. There are many such recent
surveys since the arrival of 8m telescopes with excellent echelle
spectrographs, with many more due to be published soon. 
These range from studies of the most extreme metal-poor stars, through
the `halo', `thick disk' and `thin disk' metallicity ranges, with only
the Bulge and inner disk still awaiting useful statistics. The bulge
and inner disk studies are especially needed, since the common
(modellers') presumption that `the average star forms in a disk, and
then merges into a spheroid' merits a check.

The results from these studies are outstanding science, and provide
information which must be considerde fully in any discussion of Galaxy
formation and evolution.

The common feature to all studies is the extremely well-defined
pattern of element ratios, spanning huge dynamic ranges. One example
is shown in Figure 2, taken from a superb study by Gratton etal 2003.
This figure shows the common result: different stellar population show
smooth systematic element ratio patterns across very wide dynamic ranges.

The systematic chemical abundance patterns
evident here, and the different systematics between different stellar
populations, provide undisputable evidence that the formation of the
field stars which are today in the Galactic halo happened in a single
environment, or a range of environments, in which the rate of star
formation, the rate of mass loss and the rate of chemical enrichment
had an extremely small range. Noone has yet suggested how this is
possible if indeed the halo is built up over time from a large number
of systems with uncorrelated histories.

Other recent studies extend this conclusion to the stars of the thick
disk and the old thin disk.

\begin{figure}[!ht]
\plotone{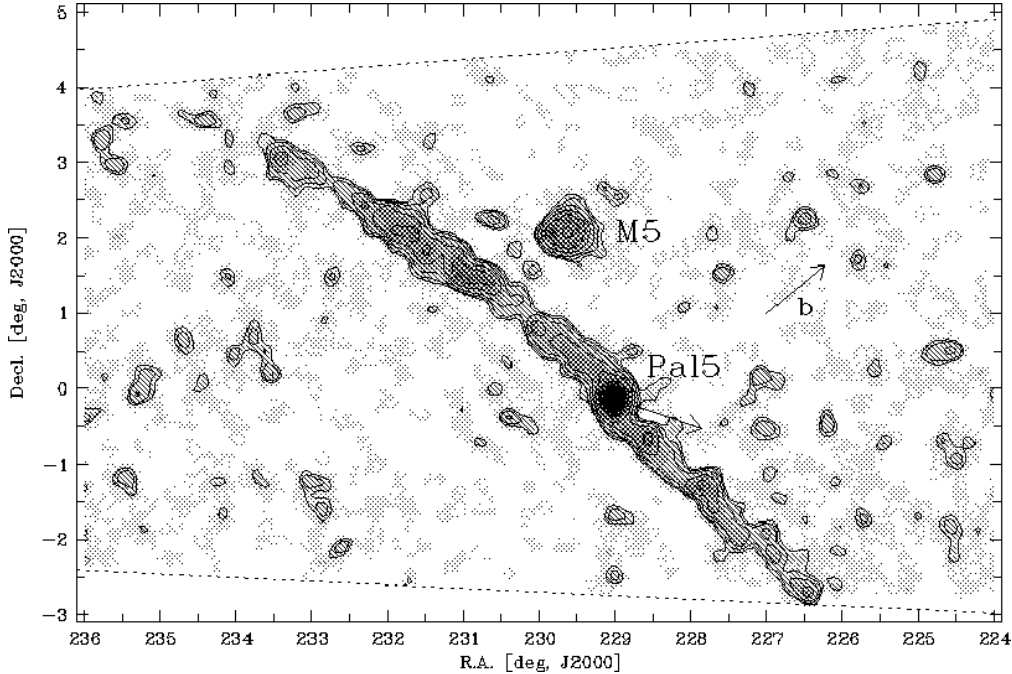}
\caption{The dissolving globular cluster Pal5 is creating phase space
structure in the Galactic halo which is unrelated to merger
histories. Some hundreds of similar phase-space structures are
expected to exist, all equally unrelated to mergers.}
\end{figure}

\section{Merger rates}

Mergers happen. The Hubble-Toomre sequence is built that way.

The Sagittarius dwarf is direct evidence. In M32 the debris from the
stripped elliptical M32 has recently been observed directly, decades
after its prediction. The Magellanic Stream exists. It is possible
that a further debris remnant has been discovered at low latitudes,
although at the time of writing the `normal' complex structure of an
outer warped disk is a more likely explanation of the observations.
Similarly, speculations that structures, such as clusters, could be
trapped into the Galactic disk from a !merger parent' need to treated
with some caution. The depth of a potential scales as the square of
its velocity dispersion, Thus the Galactic halo potential (dispersion
$\sim$100km/s) is an order of magnitude deeper than that of the disk
(dispersion $\sim$30km/s). Trapping something into the outer disk is
not an obvious dynamical process without very special geometries, and
adroit use of dynamical friction.

Is Sgr all there is? 

Phase space structures are not an indication of mergers. The most
dramatic local phase-space structures are spiral arms, Gould's belt,
and star-formation regions. Even in the halo, very considerable
phase-space structure must exist even if there has never been a single
merger in the Hubble time. 

Globular clusters evolve, and evaporate eventually. It is often
speculated that many hundreds of dissolved globular clusters exist in
the halo. A globular cluster, Pal5, has recently been observed in the
final stages of dissolution (Odenkirchen etal 2003) and is shown in
Figure 3.

The same limits on accretion histories which can be deduced from the
difference of the stellar populations of Sgr, and are quantified for
field stars by Unavane, Wyse and Gilmore (1996), can be applied to
other tracers. The severe difficulty of hiding stars with different
enrichment histories in the element ratio data is noted above.
One may do the same thing for globular clusters. Bellazzini and
collaborators are extending the census of former members of Sgr
utilising that fact that Sgr stellar populations are different from
field stars. One may do the same slightly more generally.

\begin{figure}[!ht]
\plottwo{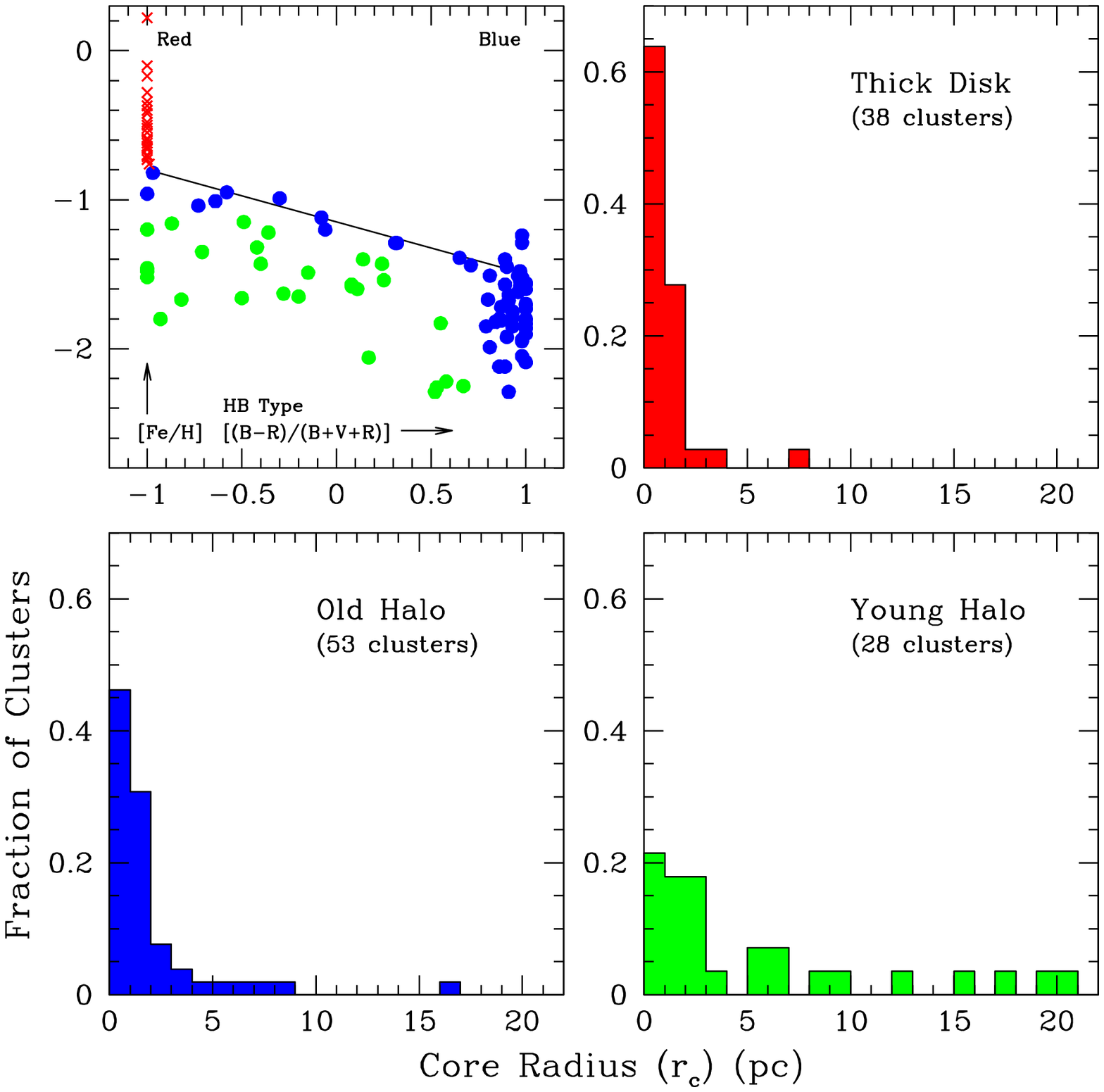}{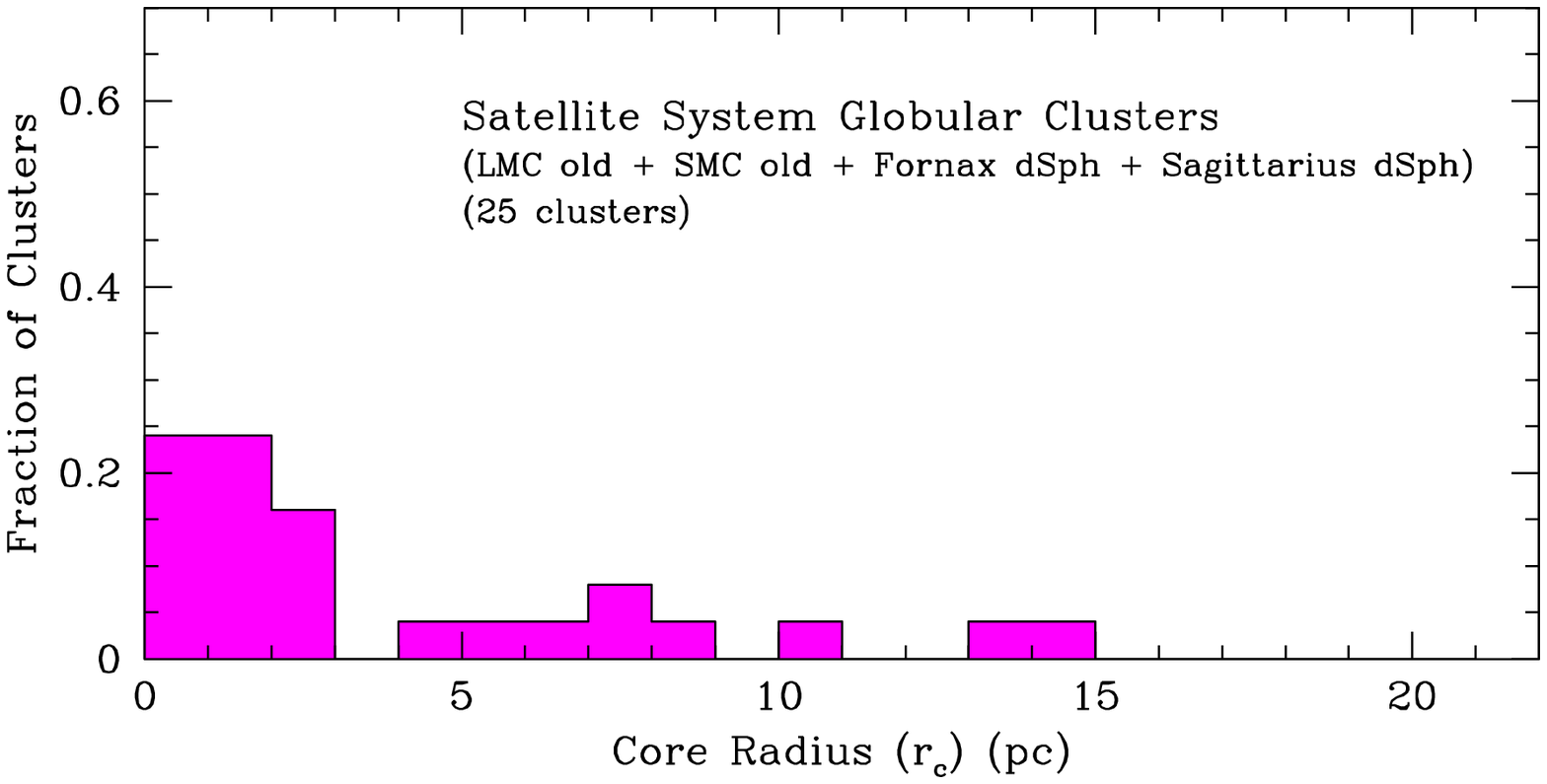}
\caption{The size distributions of the various population types of
Galactic globular clusters (LHS) and the old clusters in Galactic
satellites (RHS). These data are from Mackey \& Gilmore 2003a,2003b,2003c.}
\end{figure}

Figure 4 shows the size distribution (core radii, from of the
different groups of Galactic globular clusters, and those old clsuters
in Galactic satellites. Most clusters in Galactic satellites are
young, thereby trivially excluding any (non-Sgr) mergers in the last
$\sim$6-8Gyr with anything resembling anything which exists in the
Local group now, or its precursors. Even older mergers are interesting
in this context: the only Galactic globulars which are consistent, on
size grounds, with having arrived in a merger are the few younger
ones, including those associated already with Sgr.

Once again, we conclude that mergers with sufficient mass to perturb
the history of the Galaxy in an observable way must be rare in recent
times.

\begin{figure}[!ht]
\plotfiddle{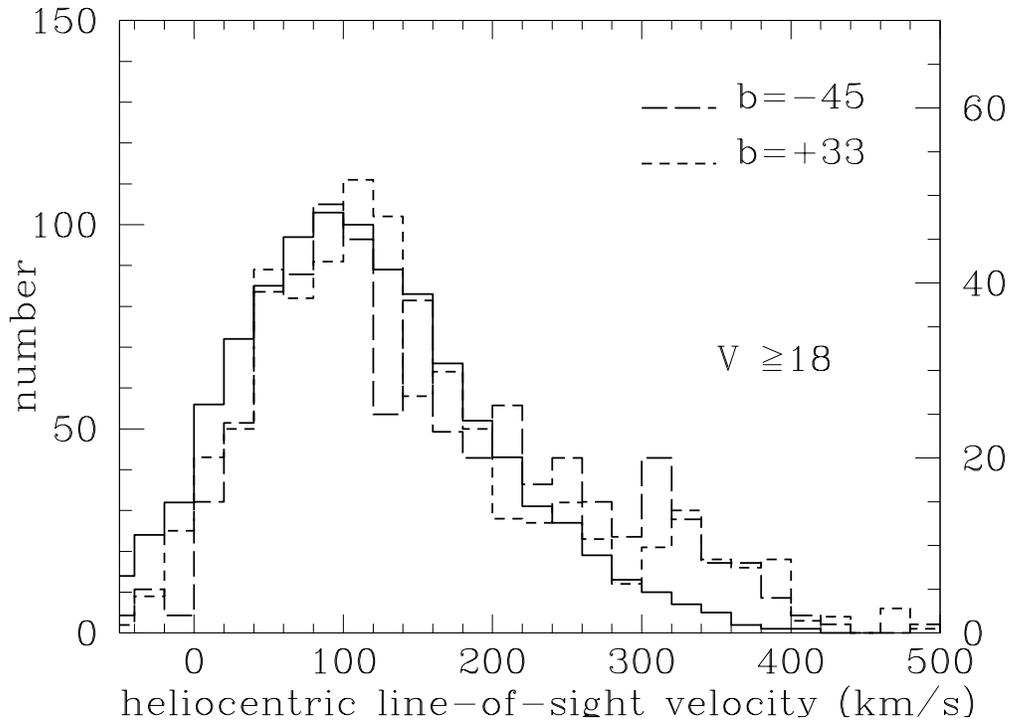}{14truecm}{-90}{50}{50}{-200}{350}
\caption{Kinematics of the thick disk in situ. The mean rotation
velocity about the Galaxy, V$\sim$100km/s
of field stars is significantly less than expected from local studies,
indicating perhaps that the (merger?) origins of the thick disk are
amenable to direct study. This figure is from Gilmore, Wyse \& Norris 2002.}
\end{figure}

\section{Field star kinematic surveys: much still to learn}

Kinematic surveys of field stars in the Galaxy remain as rare as
is evidence for multiple mergers in the Milky Way. Each of the few
attempted has made significant discoveries (eg SDSS) which should be
an encouragement. One for which the first results have recently been
reported indicates the discovery potential. Gilmore, Wyse and Norris
(2002) report first results from a survey using 2dF on the AAT. their
first results are summarised in Figure 5, and show significant
disagreement with expectation.

The mean star a few kpc from the Plane has angular momentum
intermediate between that of the halo and the thin disk. This argues
strongly for an independent origin, and is consistent with - but does
not require - a merger origin.

\section{Conclusion}

Mergers happen. The Hubble-Toomre sequence is built that way (Figure 6).

\begin{figure}
\caption{The Hubble-Toomre sequence: mergers matter, at least in
life-changing situations. But how often and when in daily existence is
less clear.  {\bf This figure is a .jpg file}}
\end{figure}

This is an age of surveys. We are learning to process and distribute
vast data volumes. We are learning to extract astrophysics. While
fashions still reign in interpretation, the story in the data is there
to be read: it tells the story of Galaxy formation.

\end{document}